\documentclass[onecolumn, draft, 12pt]{IEEEtran}
\usepackage{amsmath,cases}
\usepackage{amsthm}
\usepackage{amssymb}
\usepackage{cite}
\usepackage{amsfonts}
\usepackage{enumerate}
\usepackage[final]{graphicx}
\usepackage{gensymb}

\newcommand{\SaS}[4]{\ensuremath{S_{#1}\left(#2,\, #3,\, #4 \right)}}

\newcommand{\D}[1]{\ensuremath{\;{\rm d}#1}}

\newtheorem{theorem}{Theorem}
\newtheorem{prop}{Property}

\makeatletter

\newcommand{\Rmnum}[1]{\expandafter\@slowromancap\romannumeral #1@}
\newcommand{\decision}[2]{\overset{\mathrm{#1}}{\underset{\mathrm{#2}}{\gtrless}}}
\makeatother 

\begin{document}

\title{Distributed Detection in Coexisting Large-scale Sensor Networks}
\author{Junghoon Lee and Cihan Tepedelenlio\u{g}lu, \emph{Member, IEEE}
\thanks{The authors are with the School of Electrical, Computer, and Energy Engineering, Arizona
State University, Tempe, AZ 85287, USA. (Email:
\{junghoon.lee,cihan\}@asu.edu).} }
\maketitle

\begin{abstract}
This paper considers signal detection in coexisting wireless sensor networks (WSNs). We characterize the aggregate signal and interference from a Poisson random field of nodes and define a binary hypothesis testing problem to detect a signal in the presence of interference. For the testing problem, we introduce the maximum likelihood (ML) detector and simpler alternatives. The proposed mixed-fractional lower order moment (FLOM) detector is computationally simple and close to the ML performance, and robust to estimation errors in system parameters. We also derived asymptotic theoretical performances for the proposed simple detectors. Monte-Carlo simulations are used to supplement our analytical results and compare the performance of the receivers.

\end{abstract}


%Paper Body
\section{Introduction}
A homogeneous Poisson point process (PPP) is widely used in the literature to model large-scale wireless sensor networks (WSNs) \cite{Sung2005, Anandkumar2008, Avidor2008}. It is a natural model to use when little is known about the spatial distribution of the network, aside from the area it is spread over, and the spatial density of nodes. The complete spatial randomness or independence property makes the PPP easy to analyze. A number of studies about network interference have been reported in the literature when the interfering nodes are scattered according to a spatial PPP. In \cite{Sousa1992, Ilow1998, Yang2003, Hughes2000, Haenggi2009, Win2009}, a multi-user network is considered with power-law path loss, where the multiple access interference follows a symmetric $\alpha$-stable distribution.

As unlicensed band utilization increases, it becomes important to understand how different wireless services, operating in the same band, may affect each other. In an effort to understand the effect of coexistence between networks, the network performance was investigated in the presence of interference from a coexisting network in \cite{Howitt2001}. Coexistence issues such as congestion control and interference between WSNs and other wireless applications have been studied in \cite{Howitt2003, Cavalcanti2008}. In \cite{Nethi2011} the authors proposed a multiple access control (MAC) scheme to avoid interference between coexisting WSNs. Most WSNs may not have a sophisticated MAC scheme which requires high complexity and exchange of information. Consequently when two WSNs coexist in the same unlicensed band, the WSNs interfere with each other.

In this paper, we consider signal detection in coexisting WSNs. The spatial distribution of sensors is modeled as a PPP. The sensors measure the phenomenon of interest and transmit their local binary decisions via wireless channel to a central site for global processing. We also assume there is a coexisting sensor network which causes interference to the desired sensor network and also is modeled as an independent PPP. As a result, we consider the detection of the aggregate sensing signal from the desired network in the presence of aggregate interference from the interfering network. To the best of our knowledge, in the literature, there is no study on the signal detection with interference in coexisting WSNs. In this paper, we show that the problem becomes a binary hypothesis testing problem of detecting $\alpha$-stable random signals in $\alpha$-stable random noise, which has not been considered in the literature. In the literature on detection in $\alpha$-stable noise, the signal has always assumed to be deterministic \cite{Zozor2006, Fu2010, Ma2010, Park2011}. We design simple and robust detectors for the $\alpha$-stable signal in the presence of $\alpha$-stable interference which arises naturally from a Poisson network interference context where the signal is also from a set of Poisson sensors.

\section{Preliminaries}
\label{Preliminaries}
We provide a brief sketch of the $\alpha$-stable distribution and related theorems. We first introduce real valued $\alpha$-stable random variables, $w \sim \SaS{\alpha}{\sigma}{\beta}{\mu}$ which has a characteristic function given by \cite{Nikias_Shao, Nolan}
\begin{equation}
\label{eqn:general_char_eqn}
\varphi(t)= \exp\left( j \mu t - |\sigma t|^{\alpha}(1-j \beta \, \mathrm{sign}(t) \, \omega(t,\alpha))\right)  ,
\end{equation}
where
\begin{align}
\omega(t,\alpha) =
\begin{cases}
 \tan \left(\frac{\pi \alpha}{2} \right) &  \alpha \neq 1 \\
 -\frac{2}{\pi}\log|t| & \alpha = 1
\end{cases}
\; \; ,
\end{align}
and $\mathrm{sign}(t) = 1 \text{ if } t > 0, \mathrm{sign}(t)=0 \text{ if } t = 0,$ and $\mathrm{sign}(t)=-1 \text{ if } t < 0$. Parameter $\alpha \in (0,2]$ is the characteristic exponent, $\beta \in [-1,1]$ is the skew, $\sigma \in (0,\infty)$ is the scale and $\mu \in (- \infty, \infty)$ is the shift. When $\beta = 0$, $w$ has a symmetric distribution about $\mu$. When $\beta = 0$ and $\mu = 0$, $w$ is a symmetric $\alpha$-stable random variable. When $\alpha = 2$ and $\beta = 0$, $w$ is Gaussian, which is the only $\alpha$-stable random variable with finite variance. In this paper, we will focus on $\alpha<2$. When $\sigma = 1$ and $\mu = 0$, $w$ is said to be standardized \cite[pp. 20]{Taqqu}.

The $\alpha$-stable random variables have many useful properties, a complete list of which can be found in \cite{Taqqu, Nikias_Shao}. A property which is useful in this paper is reproduced below.
\begin{prop}
\label{prop:linear_comb}
If $ x_{i} \sim \SaS{\alpha}{\sigma_{i}}{\beta_{i}}{\mu_{i}}$, $i=1,...,N$ are independent, then
$\sum_{i=1}^{N} x_{i} \sim \SaS{\alpha}{\sigma}{\beta}{\mu}$, where
$\sigma = \left( \sum\limits_{i=1}^{N} \sigma_{i}^{\alpha} \right)^{1/\alpha}$, $\beta = \left(\sum\limits_{i=1}^{N}\beta_{i}\sigma_{i}^{\alpha}\right)\left(\sum\limits_{i=1}^{N}\sigma_{i}^{\alpha} \right)^{-1}$ and $\mu = \sum\limits_{i=1}^{N} \mu_{i}$.
\end{prop}

In the following, we will introduce the characteristic function of the aggregate signal from nodes which are distributed according to a homogeneous PPP in the two-dimensional infinite plane. Let $Y$ denote the aggregate signal from the Poisson distributed network, such that
\begin{equation}
\label{eqn:aggregate_signal}
Y = \sum^{\infty}_{i=1} \frac{X_i}{r^{\delta}_i}
\end{equation}
where $X_i$ are independent and identically distributed (i.i.d.) random sensor emissions which are used to model various propagation effects such as (multipath or shadow) fading, fluctuations in transmit power; $r_i$ is the distance between the receiver and sensor $i$, and $\delta$ is a power loss exponent. The characteristic function of $Y$ follows immediately from Campbell's theorem \cite{Kingman72} and is given by
\begin{equation}
\label{eqn:Campbell_theorem}
\varphi_{Y}(t) = \exp\left(-2\pi\lambda \int_{0}^{\infty} \left[ 1-\varphi_{X}\left(\frac{t}{r^{\delta}}\right) \right] r\mathrm{d}r\right)
\end{equation}
where $\lambda$ is the spatial density of nodes and $\varphi_{X}(\cdot)$ is the characteristic function of $X_i$. The following theorem which follows from \eqref{eqn:Campbell_theorem} gives the characteristic function of the aggregate signal $Y$ and depends on characteristic functions of random variables $X_i$.
\begin{theorem}
\label{the:symmetric_alpha}
When $\lbrace X_i \rbrace_{i=1}^{\infty}$ are i.i.d. sequence of symmetric random variables, the characteristic function of the aggregate signal $Y$ in \eqref{eqn:aggregate_signal} is given by
\begin{equation}
\label{eqn:symmetric_alpha}
\varphi_Y(t) = \exp\left(-\sigma^{\alpha}\lvert t \rvert^{\alpha}\right)
\end{equation}
where $\alpha = \frac{2}{\delta}$, $\sigma^{\alpha} = \lambda \frac{\pi}{2} C_{\alpha}^{-1} E\left[ \lvert X_i \rvert^{\alpha} \right]$ and $C_{\alpha}=\frac{1-\alpha}{\Gamma(2-\alpha)2\cos(\pi\alpha/2)}$.
\end{theorem}
\begin{proof}
See \cite{Win2009}.
\end{proof}
Random variables with characteristic function of the \eqref{eqn:symmetric_alpha} belong to the class of \emph{symmetric $\alpha$-stable random variables}. In the following, we consider the case that $\lbrace X_i \rbrace_{i=1}^{\infty}$ is a sequence of real nonnegative random variables.
\begin{theorem}
\label{the:skewed_alpha}
When $\lbrace X_i \rbrace_{i=1}^{\infty}$ are i.i.d. sequence of real nonnegative random variables, the characteristic function of the aggregate signal $Y$ in \eqref{eqn:aggregate_signal} is given by
\begin{equation}
\label{eqn:skewed_alpha}
\varphi_Y(t) = \exp\left(-\sigma^{\alpha}\lvert t \rvert^{\alpha}\left[1-j\beta\mathrm{sign}(t)\tan\left(\frac{\pi\alpha}{2}\right)\right]\right)
\end{equation}
where $\alpha = \frac{2}{\delta}$, $\beta = 1$, $\sigma^{\alpha} = \lambda \frac{\pi}{2} C_{\alpha}^{-1} E\left[ X_{i}^{\alpha} \right]$ and $C_{\alpha}=\frac{1-\alpha}{\Gamma(2-\alpha)2\cos(\pi\alpha/2)}$.
\end{theorem}
\begin{proof}
See \cite{Win2009}.
\end{proof}
Random variables with characteristic function of the form of $\varphi_Y(t)$ in \eqref{eqn:skewed_alpha} belong to the class of \emph{totally positive skewed $\alpha$-stable random variables}. Note that when $\beta=1$, each realization of $Y$ is positive.

\section{System Model}
\label{sec:System_Model}
\begin{figure}[tb]
\begin{minipage}{1\textwidth}
\centering
\begin{center}
\includegraphics[height=8.5cm,keepaspectratio]{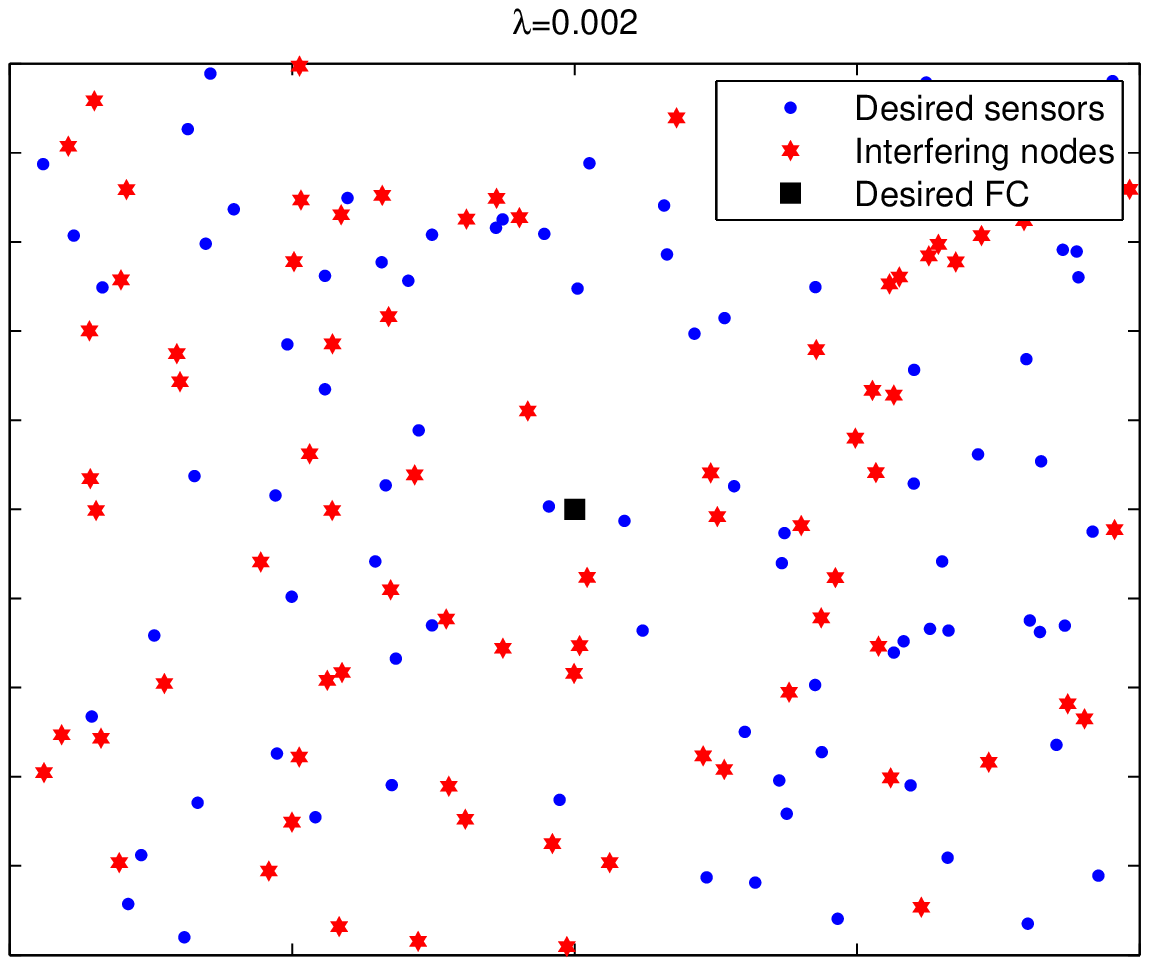}
\caption{Realization of the spatial distribution of two networks according to the homogeneous Poisson point process ($\lambda = 0.002$)}\label{fig:two_Poisson_networks}
\end{center}
\end{minipage}
\end{figure}

We consider the coexistence of two WSNs which are distributed according to PPP as shown in Fig. \ref{fig:two_Poisson_networks}. We assume one is a desired network where active sensors transmit their measurements to the Fusion Center (FC), and the other is a interfering network which causes interference to the desired network. The interfering network also may have a FC which is not shown in Fig. \ref{fig:two_Poisson_networks}. It is also assumed that the spatial density of the PPP for the desired network is $\lambda_{\mathrm{D}}$ and the spatial density of interfering network is $\lambda_{\mathrm{I}}$ respectively. Under these assumptions, we will show that both aggregate signal and interference converge to $\alpha$-stable random variables with different parameters and consider the distributed detection and data fusion problem. We will also consider imperfect local detection error at each sensor and study how this error affects the detection performance.

\subsection{Decision Fusion}
\label{sec:Decision_Fusion}
In the desired sensor network, we can consider a binary hypothesis testing problem with two hypotheses $\mathcal{H}_0, \mathcal{H}_1$ where $P_0, P_1$ are their respective prior probabilities. In this paper, we assume the two hypotheses equally likely ($P_0=P_1=1/2$). Let the sensed signal at the $i^{\text{th}}$ sensor be,

\begin{equation}
\label{eqn:sensed_signal}
x_i = \theta+n_i,
\end{equation}
where $\theta=1$ under $\mathcal{H}_1$ and $\theta=0$ under $\mathcal{H}_0$, is a deterministic parameter whose presence or absence has to be detected, and $n_i$ is the noise sample at the $i^{\text{th}}$ sensor. We consider a setting where the $i^{\text{th}}$ sensor transmits its measurement using a distributed detection scheme. It is also assumed that $h_i$ is a symmetric real-valued fading channel coefficient from the $i^{\text{th}}$ sensor to the FC satisfying $\mathbb{E}[h^2]=1$ and is known at each sensor \cite{Niu2008}.

In the system, each local sensor makes a decision based on its decision rule
\begin{align}
\label{eqn:decision_rule}
\gamma(x_i) = M_i =
\begin{cases}
1 & \mathrm{under~} \mathcal{H}_1 \\
0 & \mathrm{under~} \mathcal{H}_0
\end{cases}
\; \; \text{,}
\end{align}
where the function $\gamma(\cdot)$ is the local decision rule that minimizes the error probability at the local $i^{\text{th}}$ sensor and $M_i$ is the local decision at the $i^{\text{th}}$ local sensor. Therefore, the transmitting signal at the local sensor is sent to the FC as follows:
\begin{equation}
\label{eqn:DF_scheme}
f(x_i) = h_i\gamma(x_i) \text{.}
\end{equation}
At the desired FC, the received signal at a sample time is given by
\begin{equation}
\label{eqn:coherent_ch}
Y = \sum^{\infty}_{i=1} h_i\frac{f(x_i)}{r^{\delta}_i} + W
\end{equation}
where $W$ is the interference from the interfering network which will be shown to have a symmetric $\alpha$-stable distribution as explained later in this section. By substituting \eqref{eqn:DF_scheme} in \eqref{eqn:coherent_ch}, we obtain
\begin{equation}
\label{eqn:coherent_DF_1}
Y = \sum^{\infty}_{i=1} \frac{h_i^2}{r^{\delta}_i} M_i + W
\end{equation}
where $\lbrace h_i^2 \rbrace_{i=1}^{\infty}$ is a sequence of real non-negative random variables. In \eqref{eqn:coherent_DF_1}, $h_i^2$ corresponds to $X_i$ in \eqref{eqn:aggregate_signal}. Using Theorem \ref{the:skewed_alpha} in \eqref{eqn:coherent_DF_1}, we can show that $S=\sum^{\infty}_{i=1} h_i^2 M_i/r^{\delta}_i$ converges to the totally positive skewed $\alpha$-stable distribution as follows:
\begin{equation}
\label{eqn:coherent_DF_signal}
S \sim  S_{\alpha} \left(\sigma_S,\beta_S,0 \right)
\end{equation}
where $\alpha = \frac{2}{\delta}$, $\sigma_S^{\alpha} = \lambda_{\text{D}}\frac{\pi}{2}C_{\alpha}^{-1} M^{\alpha}E[h^{2\alpha}]$, and $\beta_S=1$.

In the following, we will show the interference from the interfering network, $W$ in \eqref{eqn:coherent_DF_1}, can be modeled as a symmetric $\alpha$-stable distribution. Similar to the signal part in \eqref{eqn:coherent_ch}, we assume that the nodes in the interfering network transmit their measurements with channel information between the nodes and \textit{their} FC. Then the aggregate interference at the desired FC at a sample time is as follows:
\begin{equation}
\label{eqn:interference}
W = \sum^{\infty}_{i=1} h_i g_i\frac{w_i}{r^{\delta}_i}
\end{equation}
where $g_i$ is the channel coefficient between the $i^{\text{th}}$ interfering node and its FC, $h_i$ is the channel between the interfering node and the desired FC, and $w_i$ is the emission at the interfering node $i$. The random variable $h_ig_iw_i$ is a symmetric regardless of $w_i$. Using Theorem \ref{the:symmetric_alpha}, we can show that \eqref{eqn:interference} converges to a symmetric $\alpha$-stable distribution as follows:
\begin{equation}
\label{eqn:coherent_AF_interference}
W \sim  S_{\alpha} \left( \sigma_W,0,0 \right)
\end{equation}
where $\alpha = \frac{2}{\delta}$ and $\sigma_W^{\alpha}=\lambda_{\text{I}}\frac{\pi}{2}C_{\alpha}^{-1}E[|hgw|^{\alpha}]$.

By using Property \ref{prop:linear_comb}, we can define the binary hypothesis testing problem as a positive $\alpha$-stable random signal detection in a symmetric $\alpha$-stable random noise, which is given as follows:
\begin{align}
\label{eqn:detect_coherent_DF}
Y =
\begin{cases}
S+W \sim S_{\alpha}(\sigma_{\mathrm{H_1}}, \beta_{\mathrm{H_1}}, 0)     & \mathrm{under~} \mathcal{H}_1 \\
W \sim S_{\alpha}(\sigma_{\mathrm{H_0}}, 0, 0)                          & \mathrm{under~} \mathcal{H}_0
\end{cases}
\; \;
\end{align}
where $\sigma_{\mathrm{H_1}}=\left(\sigma_{S}^{\alpha}+\sigma_{W}^{\alpha}\right)^{1/\alpha}$, $\beta_{\mathrm{H_1}}=\frac{\left(\beta_{S}\sigma_{S}^{\alpha}\right)}{\left(\sigma_{S}^{\alpha}+\sigma_{W}^{\alpha}\right)}$ and $\sigma_{\mathrm{H_0}}=\sigma_{W}^{\alpha}$. Note that $0<\beta_{\mathrm{H_1}}<1$ since $\sigma_{S}>0$, $\sigma_{W}>0$, and $\beta_{S}=1$.

\subsection{Detection Error at a Local Sensor}
Before we address different solutions to the binary hypothesis testing in \eqref{eqn:detect_coherent_DF}, we show that detection errors at the local sensors cause only a change in the parameters of \eqref{eqn:detect_coherent_DF}. We define the error probability for detection at a local sensor as $P_{\text{L,e}}<1$. It is also assumed that the detection error at a local sensor occurs independently across sensors. The detecting error will reduce the number of sensors which transmit their detection message to the FC. Therefore the actual transmitting nodes will be a PPP with spatial density $\lambda_{\text{D,e}}=\lambda_{\text{D}}P_{\text{L,e}}$ by the thinning property of PPP \cite{Stoyan1995}. As a result, $\sigma_S^{\alpha}$ in \eqref{eqn:coherent_DF_signal} will be reduced to $\sigma_{S,\text{e}}^{\alpha} = \lambda_{\text{D,e}}\frac{\pi}{2}C_{\alpha}^{-1} M^{\alpha}E[h^{2\alpha}]$ due to the detection error at a local sensor. This is related with the following SNR definition.

The usual SNR definition using the variance of the noise process is not suitable when the additive noise is $\alpha$-stable since $\alpha$-stable random variables do not have finite variance when $\alpha<2$. Therefore, we adopt the modified SNR definition as follows \cite{Fu2010, Ma2010}:
\begin{equation}
\label{eqn:Def_modified_SNR}
\emph{SNR}_{\text{m}}=10\log_{10}\frac{\sigma_{S}^{\alpha}}{\sigma_{W}^{\alpha}} (\text{dB})
\end{equation}
where $\sigma_{S}$ is the scale parameter of the signal, $\sigma_{W}$ is the scale parameter of the noise and $\alpha$ is the characteristic exponent. Since $\sigma_{S,\text{e}}^{\alpha} < \sigma_S^{\alpha}$, the SNR defined in \eqref{eqn:Def_modified_SNR} will decrease due to the detection error at each local sensor. Consequently, the performance will be degraded due to the detection error at a local sensor, as expected.

\section{Signal Detection}
In this section, we will introduce the Maximum Likelihood (ML) detector for $\alpha$-stable signal detection in $\alpha$-stable interference which has considerable computational complexity. To surmount the complexity of the ML detector, we will also propose several simple detectors. For the signal detection, it is assumed that the desired FC observes $\lbrace Y_j \rbrace_{j=1}^{L}$ which are $L$ independent measurements across time. The proposed detectors will apply to both error-free local detection, and also capture the presence of local errors.

\subsection{Maximum Likelihood Detector}
The optimal ML detector computes the following test statistics:
\begin{equation}
T_{\text{ML}} = \sum_{j=1}^{L}\log\frac{f_{\alpha}(Y_j,\sigma_{\mathrm{H_1}},\beta_{\mathrm{H_1}},0;\mathcal{H}_1)}{f_{\alpha} (Y_j,\sigma_{\mathrm{H_0}},\beta_{\mathrm{H_0}},0;\mathcal{H}_0)} \decision{H_1}{H_0} 0
\end{equation}
where $f_{\alpha}(Y,\sigma,\beta,\mu)$ is a probability density function of $\alpha$-stable random variable whose characteristic function is given in \eqref{eqn:general_char_eqn}. Because there is no closed form expression for the PDF of $\alpha$-stable random variables, the PDF can be obtained by taking inverse Fourier transform numerically. Instead of numerical integration, we can alternatively use a lookup table for the numerical values of $\alpha$-stable random variables. Such a lookup table would have sizable memory requirements since a lookup table would contain values for each of the $\alpha$, $\sigma$, and $\beta$ values on a fine grid corresponding to the noise parameters. Therefore, the ML detector requires high computational complexity or storage. Thus, we propose several simple detectors in following sections.

\subsection{Fractional Lower Order Moments (FLOM) Detector}
\label{FLOM_detector}
Although the second-order moment of a $\alpha$-stable random variable with $0<\alpha<2$ does not exist, all moments of order less than $\alpha$ do exist and are called the \emph{Fractional Lower-Order Moments} or FLOMs. The FLOMs of a $\alpha$-stable random variable $Y \sim S_{\alpha}(\sigma, \beta, 0)$ are given by \cite{Kuruoglu2001}:
\begin{equation}
\label{eqn:FLOM_mean}
E[|Y|^p]=\frac{\Gamma\left(1-\frac{p}{\alpha}\right)}{\Gamma(1-p)\cos\left(\frac{p\pi}{2}\right)}\left(1+\beta^2\tan^2\frac{\alpha\pi}{2}\right)^{\frac{p}{2\alpha}}\cos\left(\frac{p}{\alpha}\arctan\left(\beta\tan\left(\frac{\alpha\pi}{2}\right)\right)\right)\sigma^p
\end{equation}
for $p<\alpha$. The FLOM in \eqref{eqn:FLOM_mean} is an even function of $\beta$ and increases with $|\beta|$. Now $|Y|^p$ has a finite mean from \eqref{eqn:FLOM_mean} and a finite variance as follows:
\begin{equation}
\label{eqn:FLOM_var}
\mathrm{var}[|Y|^p]=E[|Y|^{2p}]-E[|Y|^p]^2, \text{~~} p<\alpha/2 \text{,}
\end{equation}
which can be computed using \eqref{eqn:FLOM_mean}. Since $|Y|^p$ has a finite mean and a finite variance, by the central limit theorem the mean of $|Y_j|^p, j=1,2,...,L$ will be approximately normally distributed as follows:
\begin{align}
\label{eqn:hypotheses_FLOM}
Z=\frac{1}{L}\sum_{j=1}^{L}|Y_j|^p =
\begin{cases}
\mathcal{N}(\mu_{\mathrm{H_1}},\sigma_{\mathrm{G,H_1}}^2) & \mathrm{under~} \mathcal{H}_1 \\
\mathcal{N}(\mu_{\mathrm{H_0}},\sigma_{\mathrm{G,H_0}}^2) & \mathrm{under~} \mathcal{H}_0 \\
\end{cases}
\; \;
\end{align}
where $\mu_{\mathrm{H}_k}$ and $\sigma_{\mathrm{G,H}_k}^2$ are the means and variances of Gaussian random variables under hypotheses $\mathcal{H}_k,k=0,1$. The mean $\mu_{\mathrm{H}_k}$ can be calculated from \eqref{eqn:FLOM_mean}, and $\sigma_{\mathrm{G,H}_k}^2=\mathrm{var}[|Y|^p]/L$ using \eqref{eqn:FLOM_var}. The suboptimal FLOM detector computes the following test statistics:
\begin{equation}
\label{eqn:test_FLOM}
T_{\text{FLOM}}(Z) = \log\frac{\frac{1}{\sqrt{2\pi\sigma_{\mathrm{G,H_1}}^2}}\exp\left(-\frac{(Z-\mu_{\mathrm{H_1}})^2}{2\sigma_{\mathrm{G,H_1}}^2}\right)}
{\frac{1}{\sqrt{2\pi\sigma_{\mathrm{G,H_0}}^2}}\exp\left(-\frac{(Z-\mu_{\mathrm{H_0}})^2}{2\sigma_{\mathrm{G,H_0}}^2}\right)} \decision{H_1}{H_0} 0
\end{equation}
The theoretical detection performance can be approximated as
\begin{equation}
\label{eqn:theory_FLOM}
P_{\text{e,FLOM}}=\frac{1}{2}\left[ Q\left(\frac{t_{1}-\mu_{\mathrm{H_0}}}{\sigma_{\mathrm{G,H_0}}}\right)-Q\left(\frac{t_{2}-\mu_{\mathrm{H_0}}}{\sigma_{\mathrm{G,H_0}}}\right)\right]
+\frac{1}{2}\left[ 1-Q\left(\frac{t_{1}-\mu_{\mathrm{H_1}}}{\sigma_{\mathrm{G,H_1}}}\right)+Q\left(\frac{t_{2}-\mu_{\mathrm{H_1}}}{\sigma_{\mathrm{G,H_1}}}\right)\right]
\end{equation}
where $t_1$ and $t_2$ are thresholds to decide $\mathcal{H}_1$ if $t_1 \leq Z \leq t_2$ and otherwise $\mathcal{H}_0$. These thresholds can be obtained by solving the quadratic equation arising from $T_{\text{FLOM}}(Z)=0$ in \eqref{eqn:test_FLOM}. By numerical computation, it is easy to see \eqref{eqn:theory_FLOM} is an increasing function of $p$. Thus, the detection performance becomes better as $p \to 0$. The FLOM of $\alpha$-stable random variable has been used for the estimation of the parameters of $\alpha$-stable random variables \cite{Kuruoglu2001}. In this paper, the FLOM is used for a detection problem. For radar systems, the FLOM detector for the detection of deterministic signal sequence due to $\alpha$-stable clutter with $\alpha$-stable noise has been studied in \cite{Tsihrintzis1997}. However, it is different from our system which is $\alpha$-stable random signal detection in $\alpha$-stable noise. The FLOM detector is not the best detector among multiple candidates we propose in this paper, and is improved next.

\subsection{Signed-FLOM Detector}
\label{signed_FLOM_detector}
In this section, we propose a simple detector for the $\alpha$-stable random signal detection in $\alpha$-stable noise. We denote the signed $p^{\text{th}}$ power of a number $x$ by
\begin{equation}
\label{eqn:signed_FLOM_def}
x^{<p>}:=\mathrm{sign}(x)|x|^p
\end{equation}
The signed FLOMs of a $\alpha$-stable random variable can be found as follows \cite{Kuruoglu2001}:
\begin{equation}
\label{eqn:signed_FLOM_mean}
E[Y^{<p>}]=\frac{\Gamma\left(1-\frac{p}{\alpha}\right)}{\Gamma(1-p)\sin\left(\frac{p\pi}{2}\right)}\left(1+\beta^2\tan^2\frac{\alpha\pi}{2}\right)^{\frac{p}{2\alpha}}\sin\left(\frac{p}{\alpha}\arctan\left(\beta\tan\left(\frac{\alpha\pi}{2}\right)\right)\right)\sigma^p \text{.}
\end{equation}
The variance of $Y^{<p>}$ can be defined by
\begin{equation}
\label{eqn:signed_FLOM_var}
\mathrm{var}[Y^{<p>}]=E[Y^{<2p>}]-E[Y^{<p>}]^2, \text{~~} p<\alpha/2 \text{.}
\end{equation}
Using same approach with the FLOM detector, we can define the binary hypothesis test and its test statistics same as \eqref{eqn:hypotheses_FLOM} and \eqref{eqn:test_FLOM} with different means and variances which can be calculated using \eqref{eqn:signed_FLOM_mean} and \eqref{eqn:signed_FLOM_var}. The theoretical performance also can be approximated using the $Q$-function. It is numerically observed that the performance of the signed-FLOM detector is a convex unimodal function of $p$. Therefore, one can find the optimal $p$ value numerically. The signed-FLOM statistic for $\alpha$-stable random variables has also been used for parameter estimation.

\subsection{Logarithm Detector}
In this section, we propose the another simple detector. We define $\log |Y|$ as a new random variable and a following relationship.
\begin{equation}
E[|Y|^p]=E[e^{p\log|Y|}]=M_{\log|Y|}(p)
\end{equation}
where $E[e^{p\log|Y|}]$ can be regarded as the moment generating function of $\log|Y|$. Then, moment of $\log|Y|$ of any order can be obtained by
\begin{equation}
\label{eqn:moment_Log}
E[\log|Y|^k]=\frac{\mathrm{d}^k M_{\log|Y|}(0)}{\mathrm{d}p^k} \text{.}
\end{equation}
Using \eqref{eqn:moment_Log} $\log|Y|$ has a mean and a finite variance as follows \cite{Kuruoglu2001}:
\begin{equation}
\label{eqn:Log_mean}
E[\log|Y|]=-C_{\text{e}}+\frac{C_{\text{e}}}{\alpha}+\log\sigma+\frac{\log(1+\beta^2\tan(\frac{\alpha\pi}{2})^2)}{2\alpha} \text{,}
\end{equation}
\begin{equation}
\label{eqn:Log_var}
\mathrm{var}[\log|Y|]=\frac{\pi^2}{4}-\frac{\pi^2}{6}+\frac{\pi^2}{6\alpha^2}-\frac{\arctan^2(\beta\tan(\frac{\alpha\pi}{2}))}{\alpha^2}
\end{equation}
where $C_{\text{e}}$ is the Euler constant. Since $\log|Y|$ has a mean and a finite variance, using same approach with the previous two detectors we can define the binary hypothesis test and its test statistics same as \eqref{eqn:hypotheses_FLOM} and \eqref{eqn:test_FLOM} with different means and variances which can be calculated from \eqref{eqn:Log_mean} and \eqref{eqn:Log_var}. The theoretical performance also can be approximated using the $Q$-function and it does not depend on any parameter unlike FLOM-based detectors. Like previous FLOM-based methods, the log method has also been used for the estimation of the parameters of $\alpha$-stable random variables \cite{Kuruoglu2001}.

\subsection{Mixed-FLOM Detector}
Even though we proposed several simple detectors in previous sections, the detectors have moderate performances compared with the ML detector which will be shown in Section \ref{sec:Simulations}. Therefore, we propose the novel mixed-FLOM detector which combines the FLOM and the signed-FLOM detectors with better performance and without any serious increase of computational complexity. Since $|Y|^{p_1}$ and $Y^{<p_2>}$ are functions of same random variable $Y$, there is a dependence between two random variables. The covariance between $|Y|^{p_1}$ and $Y^{<p_2>}$ is given by 
\begin{equation}
\label{eqn:mixed_FLOM_covar}
\mathrm{cov}[|Y|^{p_1},Y^{<p_2>}]=E[Y^{<p_1+p_2>}]-E[|Y|^{p_1}]E[Y^{<p_2>}], \text{~~} p_1+p_2<\alpha/2 \text{.}
\end{equation}
By using these two random variables, the binary hypothesis test can be approximated as follows:
\begin{align}
\label{eqn:hypotheses_mixed_FLOM}
\mathbf{Z}=\left[\begin{array}{c}
Z_1 \\
Z_2 \\
\end{array} \right] =
\left[\begin{array}{c}
\frac{1}{L}\sum_{j=1}^{L}|Y_j|^{p_1} \\
\frac{1}{L}\sum_{j=1}^{L}Y_j^{<p_2>} \\
\end{array} \right] =
\begin{cases}
\mathcal{N}(\boldsymbol{\mu}_{\mathrm{H_1}},\mathbf{C}_{\mathrm{H_1}}) & \mathrm{under~} \mathcal{H}_1 \\
\mathcal{N}(\boldsymbol{\mu}_{\mathrm{H_0}},\mathbf{C}_{\mathrm{H_0}}) & \mathrm{under~} \mathcal{H}_0 \\
\end{cases}
\; \;
\end{align}
where $\boldsymbol{\mu}_{\mathrm{H}_k}=[\mu_{1,k}\text{ }\mu_{2,k}]^{T}, k=0,1$ can be calculated by using \eqref{eqn:FLOM_mean} and \eqref{eqn:signed_FLOM_mean}. In covariance matrix $\mathbf{C}_{\mathrm{H}_k}=[c_{11,k}\text{ }c_{12,k} ; c_{21,k}\text{ }c_{22,k}],k=0,1$, the diagonal terms are $c_{11,k}=\mathrm{var}[|Y|^{p_1}]/L$ and $c_{22,k}=\mathrm{var}[Y^{<p_2>}]/L$ which can be calculated using \eqref{eqn:FLOM_var} and \eqref{eqn:signed_FLOM_var}. The covariance between $Z_1$ and $Z_2$ is $c_{12,k}=c_{21,k}=\mathrm{cov}[|Y|^{p_1},Y^{<p_2>}]/L$ from \eqref{eqn:mixed_FLOM_covar}. The covariance can be computed by using \eqref{eqn:FLOM_mean} and \eqref{eqn:signed_FLOM_mean}. The mixed-FLOM detector computes the following test statistics:
\begin{equation}
\label{eqn:test_mixed_FLOM}
T_{\text{mixed-FLOM}}(\mathbf{Z}) = \log\frac{\frac{1}{2\pi\det|\mathbf{C}_{\mathrm{H_1}}|}\exp\left(-(\mathbf{Z}-\boldsymbol{\mu}_{\mathrm{H_1}})^{T}\mathbf{C}_{\mathrm{H_1}}^{-1}(\mathbf{Z}-\boldsymbol{\mu}_{\mathrm{H_1}})\right)}
{\frac{1}{2\pi\det|\mathbf{C}_{\mathrm{H_0}}|}\exp\left(-(\mathbf{Z}-\boldsymbol{\mu}_{\mathrm{H_0}})^{T}\mathbf{C}_{\mathrm{H_0}}^{-1}(\mathbf{Z}-\boldsymbol{\mu}_{\mathrm{H_0}})\right)} \decision{H_1}{H_0} 0
\end{equation}
The theoretical detection performance can be approximated as
\begin{equation}
\label{eqn:theory_mixed_FLOM1}
P_{\mathrm{e,mixed-FLOM}}=\frac{1}{2}\left(P_{\mathrm{e,H_1}}+P_{\mathrm{e,H_0}}\right)
\end{equation}
where
\small
\begin{equation}
\nonumber
\label{eqn:theory_mixed_FLOM2}
P_{\mathrm{e,H_0}}=\int_{s_1}^{s_2} \left[Q\left(\frac{t_1(Z_1)-\mu_{2,0}}{\sqrt{c_{22,0}}}\right)-Q\left(\frac{t_2(Z_1)-\mu_{2,0}}{\sqrt{c_{22,0}}}\right)\right]\frac{1}{\sqrt{2\pi c_{11,0}}}\exp\left(-\frac{(Z_1-\mu_{1,0})^2}{2c_{11,0}}\right)\D Z_1 \text{,}
\end{equation}
\normalsize
and
\small
\begin{equation}
\nonumber
\label{eqn:theory_mixed_FLOM3}
P_{\mathrm{e,H_1}}=\int_{s_1}^{s_2} \left[1-Q\left(\frac{t_1(Z_1)-\mu_{2,1}}{\sqrt{c_{22,1}}}\right)+Q\left(\frac{t_2(Z_1)-\mu_{2,1}}{\sqrt{c_{22,1}}}\right)\right]\frac{1}{\sqrt{2\pi c_{11,1}}}\exp\left(-\frac{(Z_1-\mu_{1,1})^2}{2c_{11,1}}\right)\D Z_1 \text{,}
\end{equation}
\normalsize
where $t_1(Z_1) = (-b_{\mathrm{t}}+\sqrt{b_{\mathrm{t}}^2-a_{\mathrm{t}}c_{\mathrm{t}}})/a_{\mathrm{t}}$ and $t_2(Z_1) = (-b_{\mathrm{t}}-\sqrt{b_{\mathrm{t}}^2-a_{\mathrm{t}}c_{\mathrm{t}}})/a_{\mathrm{t}}$ are thresholds that are functions of $Z_1$ as seen in \eqref{eqn:Variables} to decide $\mathcal{H}_1$ if $t_1(Z_1) \leq Z_2 \leq t_2(Z_1)$ and otherwise $\mathcal{H}_0$. In order for the thresholds $t_1(Z_1)$ and $t_2(Z_1)$ to have real values, the ranges for $Z_1$ are defined as $s_1 := (-b_{\mathrm{s}}+\sqrt{b_{\mathrm{s}}^2-a_{\mathrm{s}}c_{\mathrm{s}}})/a_{\mathrm{s}}$ and $s_2 := (-b_{\mathrm{s}}-\sqrt{b_{\mathrm{s}}^2-a_{\mathrm{s}}c_{\mathrm{s}}})/a_{\mathrm{s}}$. We have also defined
\begin{equation}
\begin{aligned}
\label{eqn:Variables}
a_{\mathrm{t}} &= [\mathbf{A}]_{2,2} \\
b_{\mathrm{t}} &= [\mathbf{A}]_{1,2}Z_1+[\mathbf{b}^T]_{1,2} \\
c_{\mathrm{t}} &= [\mathbf{A}]_{1,1}Z_1^2+2[\mathbf{b}^T]_{1,1}Z_1+c \\
a_{\mathrm{s}} &= [\mathbf{A}]_{1,2}^2-[\mathbf{A}]_{2,2}[\mathbf{A}]_{1,1} \\
b_{\mathrm{s}} &= [\mathbf{A}]_{1,2}[\mathbf{b}^T]_{1,2}-[\mathbf{A}]_{2,2}[\mathbf{b}^T]_{1,1} \\
c_{\mathrm{s}} &= [\mathbf{b}^T]_{1,2}^2-c[\mathbf{A}]_{2,2} \text{,}
\end{aligned}
\end{equation}
where $\mathbf{A}=\mathbf{C}_{\mathrm{H_0}}^{-1}-\mathbf{C}_{\mathrm{H_1}}^{-1}$, $\mathbf{b}^T=\boldsymbol{\mu}_{\mathrm{H_1}}^T\mathbf{C}_{\mathrm{H_1}}^{-1}-\boldsymbol{\mu}_{\mathrm{H_0}}^T\mathbf{C}_{\mathrm{H_0}}^{-1}$, $c=\boldsymbol{\mu}_{\mathrm{H_0}}^T\boldsymbol{\mu}_{\mathrm{H_0}}-\boldsymbol{\mu}_{\mathrm{H_1}}^T\boldsymbol{\mu}_{\mathrm{H_1}}+\log\frac{\det|\mathbf{C}_{\mathrm{H_0}}|}{\det|\mathbf{C}_{\mathrm{H_1}}|}$, and $[\mathbf{D}]_{i,j}$ is the $(i,j)^{\text{th}}$ element of matrix $\mathbf{D}$. Using \eqref{eqn:theory_mixed_FLOM1}, we can obtain optimal $p_1$ and $p_2$ values which guarantee the best theoretical performance numerically.

The mixed-FLOM detector requires the mean $\boldsymbol{\mu}_{\mathrm{H}_k}$ and covariance matrix $\mathbf{C}_{\mathrm{H}_k}$, $k=0,1$ which can be calculated using \eqref{eqn:FLOM_mean}, \eqref{eqn:FLOM_var}, \eqref{eqn:signed_FLOM_mean}, and \eqref{eqn:signed_FLOM_var} with the parameters of $\alpha$-stable random variable. The mixed-FLOM detector has close performance to the ML detector and is more robust to uncertainties in the knowledge of parameters of $\alpha$-stable random variables than the ML detector which will be shown in the next Section \ref{sec:Simulations}.

\section{Simulations}
\label{sec:Simulations}
In this section, we will show the detection performances for the proposed detectors. Since signal and interference models are already verified mathematically in Section \ref{sec:System_Model}, we generate totally positive skewed $\alpha$-stable random variables and symmetric $\alpha$-stable random variables as the signal and noise instead of aggregate signal and interference from Poisson networks.

\begin{figure}[tb]
\begin{minipage}{1\textwidth}
\centering
\begin{center}
\includegraphics[height=8.5cm,keepaspectratio]{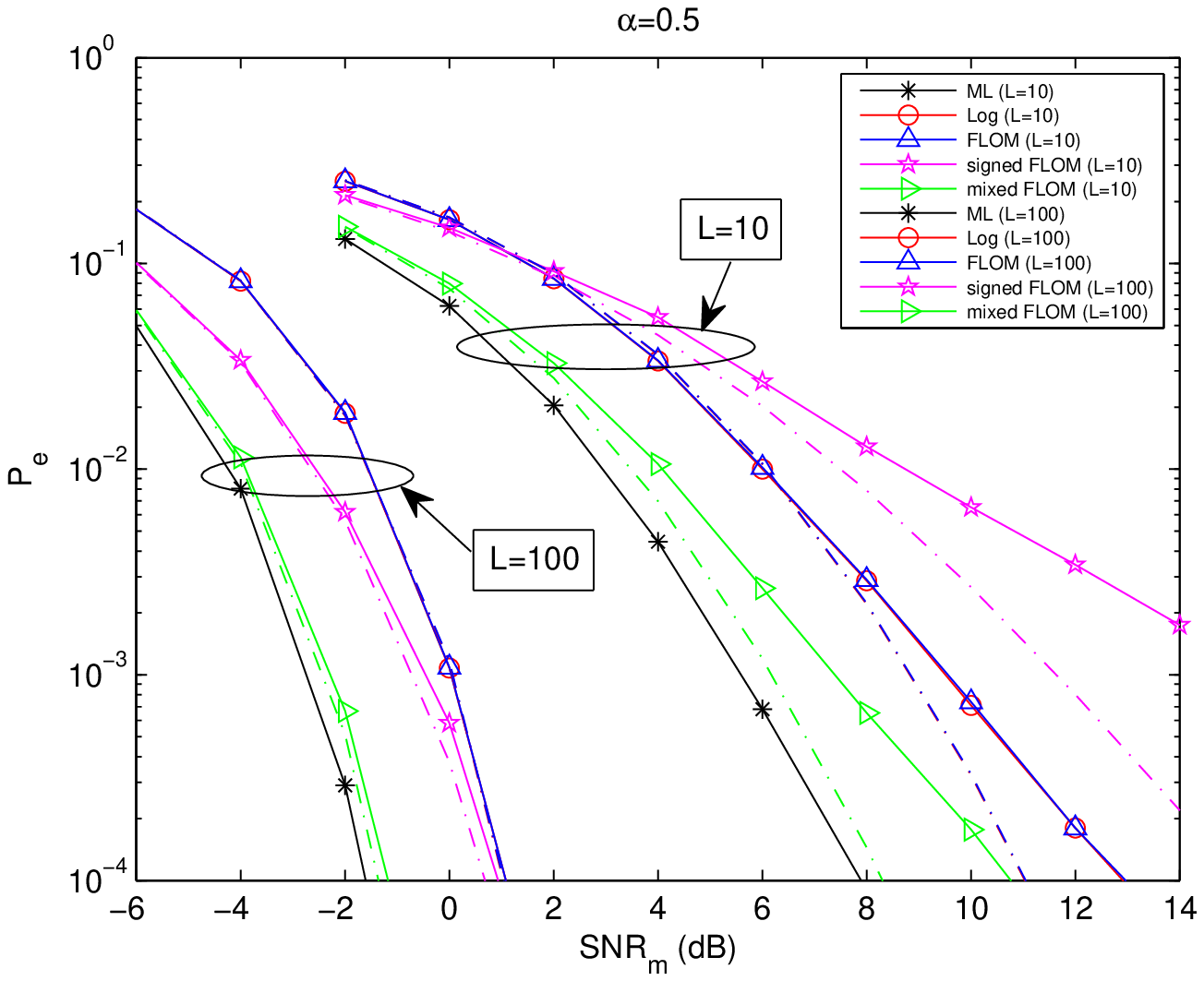}
\caption{Performance comparison of ML and proposed simple detectors with $\alpha=0.5$. The dotted lines are the theoretical results using \eqref{eqn:theory_FLOM} and \eqref{eqn:theory_mixed_FLOM1} with different means and variances according to the proposed detectors.}\label{fig:Alpha_0p5}
\end{center}
\end{minipage}
\end{figure}

In Fig. \ref{fig:Alpha_0p5}, we show the detection performance of proposed detectors with $\alpha=0.5$ which implies the path-loss exponent is $\delta=4$. We also show the performance of the ML detector for comparison using numerical integration. The number of received time samples for detection is $L$. For the FLOM detector, $p=0.001$ is used since the theoretical performance of the FLOM detector is an increasing function of $p$. In case of the signed-FLOM detector, we used the optimal $p$ values which are obtained numerically and guarantee the minimum theoretical performance. The log detector does not depend on $p$. The simulated performances of proposed detectors come close to their theoretical performances as $L$ increases due to the central limit theorem. The mixed-FLOM detector shows the best performance over other proposed detectors. When the number of samples for detection is small ($L=10$), the gap between the mixed-FLOM detector and the ML detector is about 2 dB at $10^{-3}$ error rate. Meanwhile, the performance gaps between the ML and the other proposed detectors are more than 3 dB. The gap between the ML and the mixed-FLOM detector decreases to less than 0.5 dB when the number of samples for detection is relatively large ($L=100$).

\begin{figure}[tb]
\begin{minipage}{1\textwidth}
\centering
\begin{center}
\includegraphics[height=8.5cm,keepaspectratio]{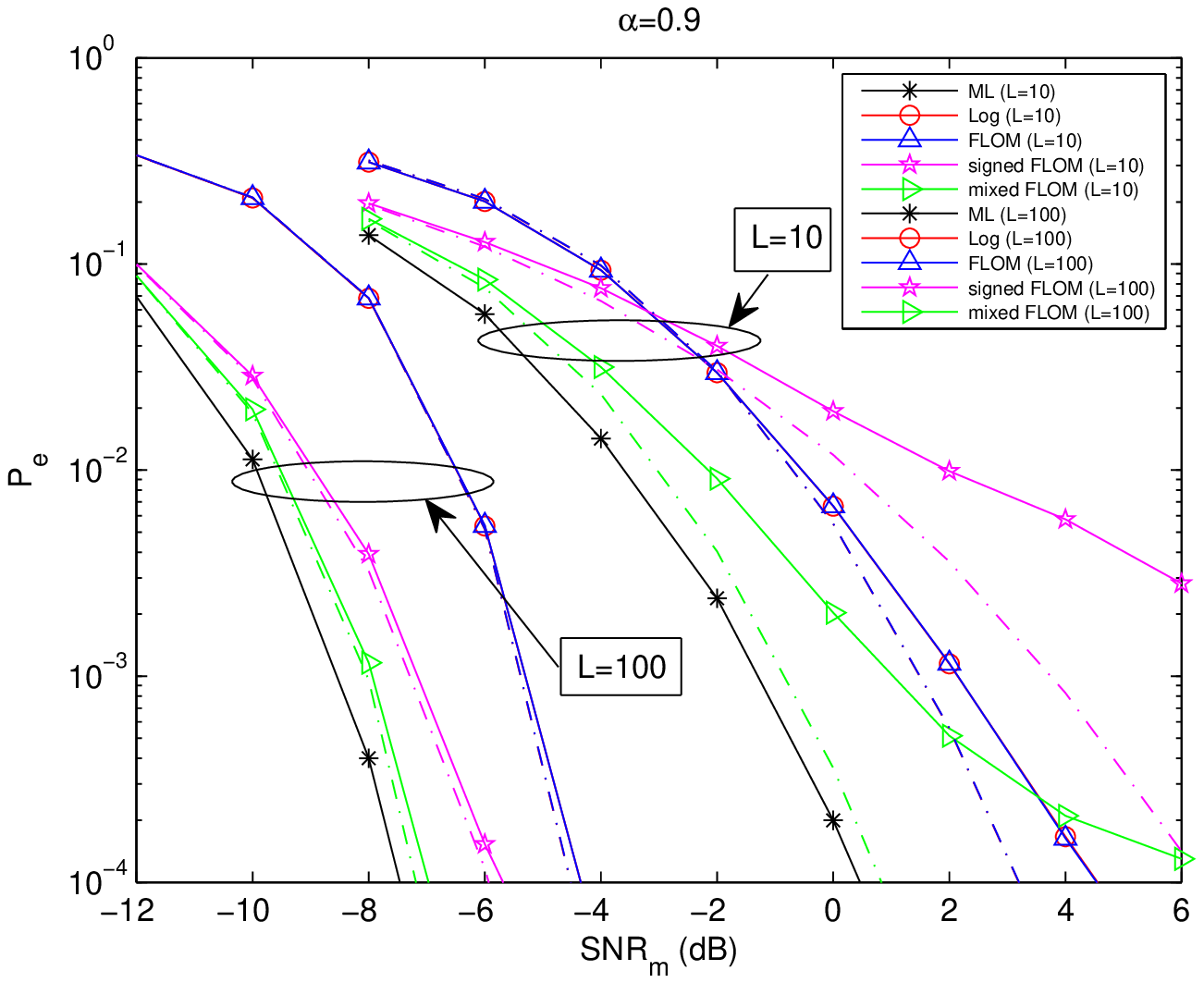}
\caption{Performance comparison of ML and proposed simple detectors with $\alpha=0.9$. The dotted lines are the theoretical results using \eqref{eqn:theory_FLOM} and \eqref{eqn:theory_mixed_FLOM1} with different means and variances according to the proposed detectors.}\label{fig:Alpha_0p9}
\end{center}
\end{minipage}
\end{figure}

In the following, we show the performances with $\alpha=0.9$ corresponding to the path-loss exponent $\delta=2.222$ which is close to the path-loss exponent in free space. In Fig. \ref{fig:Alpha_0p9}, the mixed-FLOM detector is seen to be within 0.7 dB of the ML detector at $10^{-3}$ error rate when the number of samples for detection is sufficiently large ($L=100$).

\begin{figure}[tb]
\begin{minipage}{1\textwidth}
\centering
\begin{center}
\includegraphics[height=8.5cm,keepaspectratio]{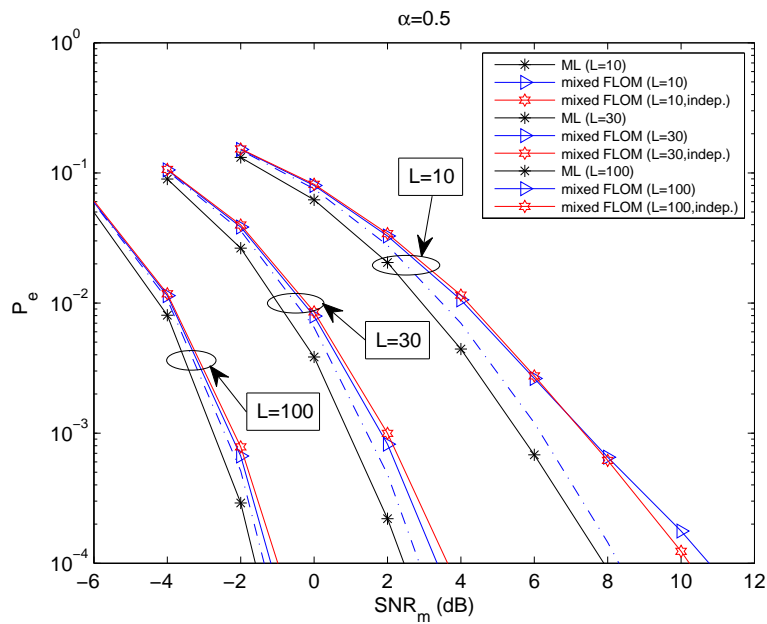}
\caption{Performance comparison of mixed-FLOM detector with its simply modified detector with $\alpha=0.5$ The dotted lines are the theoretical results using \eqref{eqn:theory_mixed_FLOM1}.}\label{fig:Alpha_0p5_indep}
\end{center}
\end{minipage}
\end{figure}

Even though the complexity of mixed-FLOM detector is much less than the ML detector, it still requires the inverse of a $2\times2$ covariance matrix \eqref{eqn:test_mixed_FLOM}. If it is assumed that $Z_1$ and $Z_2$ are independent in \eqref{eqn:hypotheses_mixed_FLOM}, the mixed-FLOM detector does not require the inverse matrix operation with negligible performance loss as shown in Fig. \ref{fig:Alpha_0p5_indep}. It is noted that the mixed-FLOM detector with this assumption has a better performance than the mixed-FLOM detector without the assumption at high SNR regime when the small number of $L$ is used for detection ($L=10$). The reason is that positive $\alpha$-stable random variables will be generated more likely than negative values under $H_1$ since $\beta_{\mathrm{H_1}}$ is close to $1$ at high SNR $\sigma_S^{\alpha}\gg\sigma_W^{\alpha}$. In this case, if $L$ is small, $Z_1$ and $Z_2$ have strong correlation even though $p_1$ and $p_2$ have different values. The strong correlation can cause the covariance matrix to be ill-conditioned. By the independence assumption between $Z_1$ and $Z_2$, the ill-conditioning can be avoided. However, except for the small number of samples for detection ($L=10$), a small loss of performances is seen due to disregarding the dependency between two random variables, $Z_1$ and $Z_2$, in Fig. \ref{fig:Alpha_0p5_indep}.

\begin{figure}[tb]
\begin{minipage}{1\textwidth}
\centering
\begin{center}
\includegraphics[height=8.5cm,keepaspectratio]{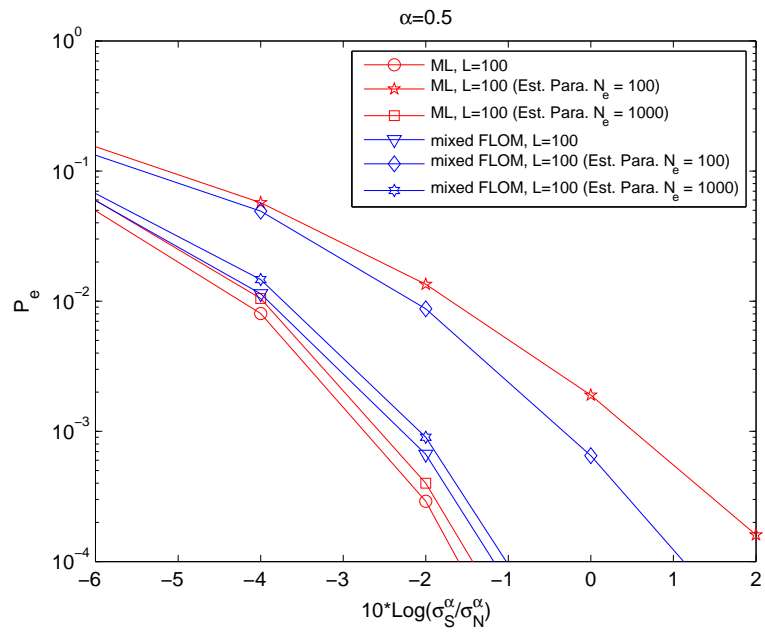}
\caption{Performance comparison of ML and mixed-FLOM detector with parameter estimation error.}\label{fig:Est_para}
\end{center}
\end{minipage}
\end{figure}

Fig. \ref{fig:Est_para} shows the performances when the detectors use the estimated parameters of $\alpha$-stable random variables, $\alpha$, $\sigma$, and $\beta$. In order to investigate the effect of parameter estimation error, we assume the parameters are estimated periodically under both hypotheses using existing estimation schemes \cite{Kuruoglu2001} and then used for the signal detection. When the number of samples for estimation is relatively large $(N_{\mathrm{e}}=1000)$, the ML detector is slightly better than the mixed-FLOM detector similar with the case that the exact knowledge of the parameters is assumed. However, the ML detector is {\it worse} than the mixed-FLOM detector when the number of samples for estimation is relatively small $(N_{\mathrm{e}}=100)$. Therefore, the mixed-FLOM detector not only takes a advantage of the low computational complexity, but also possesses robustness to the parameter estimation error.

\section{Conclusions}
\label{sec:Conclusions}
In this paper, we assumed two Poisson distributed networks coexist. In this environment, we showed the signal and interference converged to the class of $\alpha$-stable distributions. From these results we defined the $\alpha$-stable random signal detection problem in $\alpha$-stable random noise. Since the ML detector is computationally complex, we have also developed the mixed-FLOM detector, which performs within 0.5 dB to the ML detector when the number of sample for detection is not small. We verified our results through Monte Carlo simulations.

\bibliographystyle{IEEEtran}
\nocite{*}
\bibliography{references}
\newpage

\end{document}